\documentstyle[12pt,twoside,psfig]{article}
\begin{document}
\begin{center}
{\Large\bf Can neutrino viscosity drive the late time cosmic acceleration?}
\\[15mm]
Sudipta Das$^{\dagger,}$\footnote{E-mail:sudipta@hri.res.in}~~and 
Narayan Banerjee$^{\dagger\dagger,}
$\footnote{E-mail: narayan@iiserkol.ac.in}$^{,}
$\footnote{ On lien from Department of Physics, Jadavpur University, 
Kolkata 700 032, India.}\\
$^\dagger${\em Harish-Chandra Research Institute, Chhatnag Road, 
Jhunsi, Allahabad 211 019, India}\\
$^{\dagger\dagger}${\em Indian Institute of Science Education 
and Research (IISER)- Kolkata, Sector-III,\\ 
Salt Lake, Kolkata - 700 106, India.}
\end{center}

\vspace{0.5cm}
\vspace{0.5cm}

\pagestyle{myheadings}
\newcommand{\be}{\begin{equation}}
\newcommand{\ee}{\end{equation}}
\newcommand{\bea}{\begin{eqnarray}}
\newcommand{\eea}{\end{eqnarray}}
\newcommand{\bc}{\begin{center}}
\newcommand{\ec}{\end{center}}
\begin{abstract}
In this paper it has been shown that the neutrino bulk viscous stresses 
can give rise to the late time acceleration of the universe. It is found 
that a number of spatially flat FRW models with a negative deceleration 
parameter can be constructed using neutrino viscosity and one of them 
mimics a $\Lambda$CDM model. This does not require any exotic dark 
energy component or any modification of gravity.
\end{abstract}

\section{Introduction}
That the present universe is undergoing an accelerated expansion has 
now been firmly established. The initial indications came from the 
supernovae data \cite{1} and were soon confirmed by many high 
precision observations including the WMAP \cite{2}. Theoretical physics has 
thus been thrust with the challenge of finding the agent, dubbed 
``{\em dark energy}", which can drive this acceleration. Naturally 
a host of candidates appeared in the literature which can provide 
this antigravity effect. However, the dark energy should become 
dominant only during the later stages of matter era so that 
nucleosynthesis in the early universe and the large scale structure 
formation in the matter dominated regime could proceed unhindered 
and make the universe look the place where we live in now.
\par Amongst the various dark energy candidates, the cosmological 
constant $\Lambda$ is certainly the most talked about one. It matches 
different observational requirements quite efficiently and has been 
known in cosmology for quite a long time for the various roles it could 
play. The insurmountable problem is of course that of the huge discrepancy 
between the required value of $\Lambda$ and that predicted 
theoretically \cite{3}. 
\par The quintessence models, where an effective negative pressure 
generated by a scalar field potential drives the acceleration, work 
extremely well to fit into various observational constraints. For a 
very brief review, we refer to Martin's recent work \cite{4}. 
But none of the potentials employed for the purpose can boast of any 
sound theoretical motivation. Non minimally coupled scalar field theories 
like Brans - Dicke theory can also be used where even a dark energy is 
not required \cite{5}, but the value of the Brans - Dicke parameter 
$\omega$ needs to be given a very small value contrary to the 
observational requirements. Scalar - tensor theories with a dark energy, 
particularly where an interaction between the dark energy and the 
geometrical scalar field like the Brans - Dicke scalar field is 
allowed so as to alleviate the coincidence problem, appear to do 
well \cite{6}. But the nature of the interaction is hardly 
well-motivated. Also, a recent work shows that energy has to be pumped 
in to the dark matter from the dark energy sector as demanded by the 
second law of thermodynamics \cite{7}. This is indeed counter-intuitive 
in view of the fact that the dark energy dominates over the dark matter 
only during the later stages of evolution.  
Chaplygin gas models \cite{8}, modified gravity theories \cite{9} and 
many other models are proposed. They all have their success stories as 
well as failures in some way or the other. In the absence of a clear 
verdict in favour of a particular dark energy candidate, all of them 
have to be discussed seriously.  
There are excellent reviews regarding 
different models and their relative merits \cite{10}. 
\par One important general feature is that all these models use either 
some kind of an exotic field or some modification of the firmly 
established general relativity - the effect of the modification being 
 hardly required by 
other branches of well established physical theories, and the possibility of 
an actual physical detection of them appears to be quite a far-fetched one.
\par Recently a well known sector of matter, whose existence in abundance 
has been firmly established, namely the neutrino distribution, has been 
proposed as a candidate for the dark energy \cite{11}. The motivation 
comes from particle physics, and for a brief but comprehensive 
review we refer to \cite{12}. However, in these models, neutrinoes 
are normally the sector which `{\em feels}' the existence of dark 
energy and cannot really solve the problem by itself, i.e, without a 
quintessence potential. For a review of the neutrino properties in 
a cosmological context, we refer to \cite{dolgov}.
\par In the present work, we treat the neutrinoes completely classically 
and show that bulk viscous stresses in the distribution can indeed do 
the trick.  
\par The advantage of the neutrinoes is that they are real objects, 
the method of detection being quite well conceived. The neutrinoes 
were decoupled from the background radiation quite early in the evolution 
when the temperature was as high as $3 \times 10^{10} K$ \cite{13}. Thus the 
interaction of the neutrinoes with other forms of matter can be ignored and 
hence the problem of the ``direction of the flow of energy" \cite{7} does not 
arise and the model becomes much more tractable. 
\par Neutrino viscosity, both in the form of shear \cite{14} or in 
the form of bulk viscosity \cite{15} had been investigated quite a 
long time back for various purpose. There has been a renewed 
interest in the neutrino viscosity quite recently as well \cite{16}. 
The problem of dissipative effects like viscosity or heat 
conduction had been that of a parabolic transport equation, 
which could allow the 
signals to travel with super-luminal speed resulting in a violation of 
causality. But the extended irreversible thermodynamics, which modifies 
the transport equation by including a relaxation time and a further 
divergence term to avoid this problem, is now quite well understood 
\cite{17} and the modified transport equation had already been 
quite extensively used in cosmology \cite{18}. 
\par In what follows, we employ a two - component non-interacting matter 
sector, one is the normal cold dark matter and the other being a 
neutrino distribution. The latter is endowed with bulk viscosity which 
produces a negative stress. It is shown that a very simple 
accelerated model can be constructed from this. The model looks simple as  
the shear viscosity is neglected in order to be consistent with the isotropic 
nature of the universe.
\section{Field equations and results :~}
\begin{figure}[!h]
\centerline{\psfig{figure=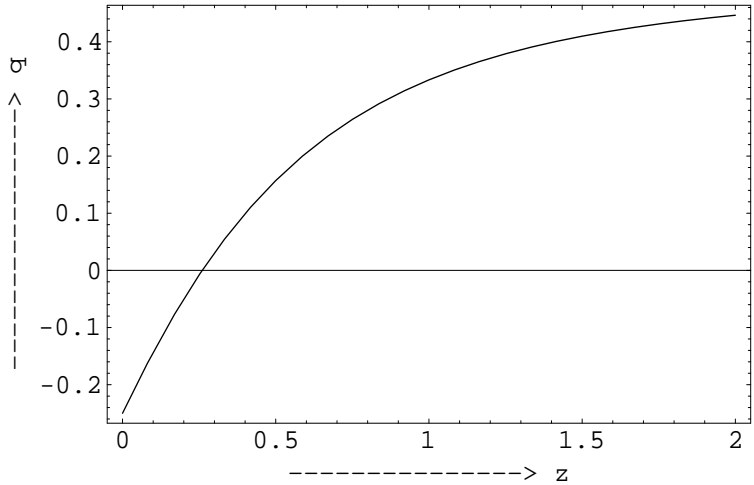,height=65mm,width=100mm}}
\caption{\normalsize{\em Plot of $q$ vs. $z$ for the choice of $a$ given by 
equation (\ref{sf}).}}
\label{fig1}
\end{figure}

Einstein equations for a spatially flat FRW universe are given by 
\be\label{fe1}
3\frac{{\dot{a}}^2}{a^2} = \rho_{m} + \rho_{n}~,
\ee
\be\label{fe2}
2\frac{\ddot{a}}{a} + \frac{{\dot{a}}^2}{a^2} = - \Pi~,
\ee
where $a$ is the scale factor for the model given by 
\bc
$ds^2 = dt^2 - a^2(t)(dr^2 + r^2 d\Omega^2)$~;
\ec
$\rho_{m}$ and $\rho_{n}$ are the densities of the normal matter and the 
neutrino distribution respectively and an overhead dot represents a 
differentiation with respect to the cosmic time $t$. Consistent with 
the present universe, we assume that both the forms of matter are 
pressureless. The advent of massive neutrinoes provides the possibility that
they can be non-relativistic so that the pressure can be neglected as in
the case of the cold dark matter. The equations are written in 
units where $8\pi G = 1$. 
The bulk viscous stress satisfies the causal transport equation
\be\label{transeq} 
\Pi + \tau \dot{\Pi} = -3 \eta H - \frac{1}{2}\tau \Pi \left[3 H + 
       \frac{\dot{\tau}}{\tau} - \frac{\dot{\eta}}{\eta} - 
                \frac{\dot{T}}{T}\right]~,
\ee
where $\tau$ is the relaxation time for the dissipative effects, $\eta$ 
is the co-efficient of viscosity and $T$ is the temperature. Being 
non-interacting amongst each other, both $\rho_{m}$ and $\rho_{n}$ 
satisfy their own conservation equations. For $\rho_{m}$, the 
conservation equation 
\bc
$\dot{\rho_{m}} + 3 H \rho_{m} = 0$
\ec
immediately integrates to yield 
\be\label{matterconsv}
\rho_{m} = \frac{\rho_{m0}}{a^3}~,
\ee
$\rho_{m0}$ being a constant of integration. Equations (\ref{fe1}), 
(\ref{fe2}) and (\ref{matterconsv}) combine to give the conservation 
equation for $\rho_{n}$ as 
\be\label{neutrinoconsv}
\dot{\rho_{n}} + 3 H (\rho_{n} + \Pi) = 0~,
\ee
which indeed is not an independent equation. So, one has three 
unknowns, namely $a$, $\rho_{n}$ and $\Pi$ whereas only two equations 
to solve for them. In order to close the system of equations, another 
equation will be required. As we require a specific dynamics for 
the universe, which is accelerating at the present moment but had 
experienced a more sedate form of a decelerated expansion during a 
none - too - distant past in the matter dominated regime itself, 
we take the form of $a$ as 
\be\label{sf}
a = a_{0} {[\mathrm{sinh} (\alpha t)]}^{2/3}~,
\ee
as the third equation. Here $a_{0}$ and $\alpha$ are positive constants. 
This simple ansatz gives the required behaviour. For small values of `$t$', 
\bc
$a \sim t^{2/3}$~,
\ec
which indeed yields a decelerated expansion. In fact the deceleration 
parameter 
\bc
$q = -\frac{\ddot{a}/a}{{\dot{a}}^2/a^2} = \frac{1}{2}$~,
\ec
which is positive and exactly same as that for a matter dominated 
universe without any other field. For a large $t$, 
\bc
$a \sim e^{2t/3}$~.
\ec
The universe is exponentially expanding and hence accelerated with $q = -1$. 
The evolution of $q$ with the redshift parameter $z$ for 
the choice of $a$ as in equation 
(\ref{sf}) is given in figure \ref{fig1} which indicates that $q$ has a smooth 
transition from a positive to a negative phase. As $z$ is given by 
\bc
$1 + z = \frac{a_{0}}{a}$~,
\ec
where $a_{0}$ is the present value of $a$, $z$ is a measure of the epoch 
in the past that one is looking at. The present value of $z$ is zero.  
\par Equation (\ref{sf}) can be used in equation (\ref{fe2}) to find 
$\Pi$ as 
\be
\Pi = -\frac{4 {\alpha}^2}{3}~,
\ee
which is a constant. Equation (\ref{neutrinoconsv}) can now be integrated 
to yield 
\be 
\rho_{n} = \frac{\rho_{n0}}{a^3} + \frac{4 {\alpha}^2}{3}~,
\ee
where $\rho_{n0}$ is a constant of integration. So, now the model is 
completely solved, it allows for the observed behaviour of $q$ and is 
extremely simple.
\par The great advantage of this model is the constancy of $\Pi$. This 
effectively gives rise to a $\Lambda$CDM model, as Einstein's equations 
now look like 
\be\label{fe1new}
3\frac{{\dot{a}}^2}{a^2} = \frac{\rho_{m0} + \rho_{n0}}{a^3} 
         + \frac{4 {\alpha}^2}{3}~,
\ee
\be\label{fe2new}
2\frac{\ddot{a}}{a} + \frac{{\dot{a}}^2}{a^2} = \frac{4 {\alpha}^2}{3}~.
\ee
The constant value $\frac{4 {\alpha}^2}{3}$ of $\Pi$ serves the purpose 
of a cosmological constant. So one can have all the virtues of a 
$\Lambda$CDM model, without having to write a cosmological constant 
by hand. An effective $\Lambda$ is provided by the bulk viscous stresses 
of neutrinoes, which is already quite well understood. So the model does 
not require any exotic field or a modification of gravity.  
\par The high degree of non linearity of Einstein's equations allows 
us to have other solutions as well. So this beautiful solution is 
by no means a unique one. In fact there could be other consistent 
accelerated solutions for equations (\ref{fe1}), (\ref{fe2}) and 
(\ref{neutrinoconsv}). One example is 
\be\label{sf2} 
H = \frac{\dot{a}}{a} = A \left(1 + a^{-3/2}\right)~,
\ee
where $A$ is a constant. This solution also describes the present 
acceleration quite efficiently as discussed by the present authors 
\cite{19}. This ansatz also satisfies all the field equations. 
However, this will not mimic a $\Lambda$CDM model as $\Pi$ comes 
out to be evolving rather than being a constant. Using equation 
(\ref{sf2}) in equation (\ref{fe2}), one can write $\Pi$ as 
\bc
$\Pi = - 3 A H$~.
\ec
\par
\begin{figure}[!h]
\centerline{\psfig{figure=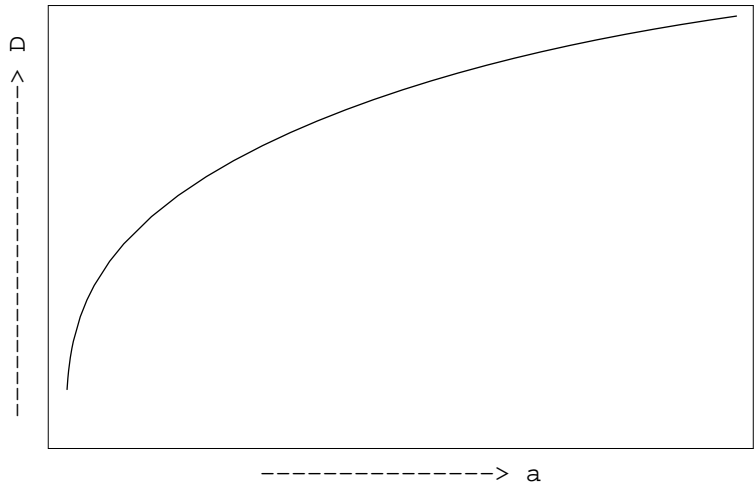,height=65mm,width=100mm}}
\caption{\normalsize{\em Plot of density contrast $D$ vs. $a$ for 
$\alpha = 1$ which clearly shows a growing mode for the dark 
matter distribution.}}
\label{fig2}
\end{figure}

 The density perturbation for the ``effectively $\Lambda$CDM'' 
model given by equation (\ref{sf}) 
can be studied. The linearized perturbation equation 
\be
\ddot{D} + 2 H \dot{D} - 4 \pi G \rho_{m} D = 0 
\ee
for the model can be numerically integrated to yield the growth of the 
density perturbation as shown in figure \ref{fig2}. The density contrast $D$ is 
as usual given by 
\bc
$D = \frac{\rho_{m} - \bar{\rho}}{\bar{\rho}}$
\ec
where $\rho_{m}$ is the perturbed density and $\bar{\rho}$ is the background 
density. 
\par The temperature profile of the neutrino distribution can also be 
estimated with the help of the equation (\ref{transeq}). For this purpose, 
the coefficient of viscosity $\eta$ and the relaxation time $\tau$ 
have to be given in terms of quantities like the density. The popular 
choice of $\eta = \eta_{0} \rho^{m}$  and $\tau = \frac{\eta}{\rho}$ where 
$m$ is a constant \cite{17}, however, does not seem to work well. With 
these choices and for $ m = \frac{1}{2}$, equation (\ref{transeq}) 
provides an expression for the 
neutrino temperature $T$ as 
\be
T = T_{0}{\left(\frac{X^{\sqrt{p}}\left(2 \sqrt{X + 1} \sqrt{X + p} 
     + 2 X + p + 1\right)}{\left(2 \sqrt{p} \sqrt{X + 1} 
                   \sqrt{X + p} + (p + 1) X + 2 p\right)^{\sqrt{p}}}\right)
                  }^{-\frac{1}{\eta_{0} \alpha}
         \sqrt{\frac{\rho_{n0}}{{a_{0}}^3}}}
\left(\frac{X}{X + p}\right)e^{\frac{2 X}{p}}
\ee
where $X = {\mathrm{cosech}}^2(\alpha t) = \left(\frac{a_{0}}{a}\right)^3 = 
(1 + z)^3$ and 
$p = \frac{4 {\alpha}^2 {a_{0}}^3}{3 \rho_{n0}}$. The plot of $T$ 
vs. $z$ shows that the neutrino temperature given by this ansatz shoots to 
a very high value at low values of the redshift $z$ ($ z \sim 1$) ( see 
figure 3), whereas the neutrino temperature is known to be below that 
of the cosmic microwave background radiation. Some other choices of 
$\eta$ and $\tau$ may provide a better temperature profile. 
\begin{figure}[!h]
\centerline{\psfig{figure=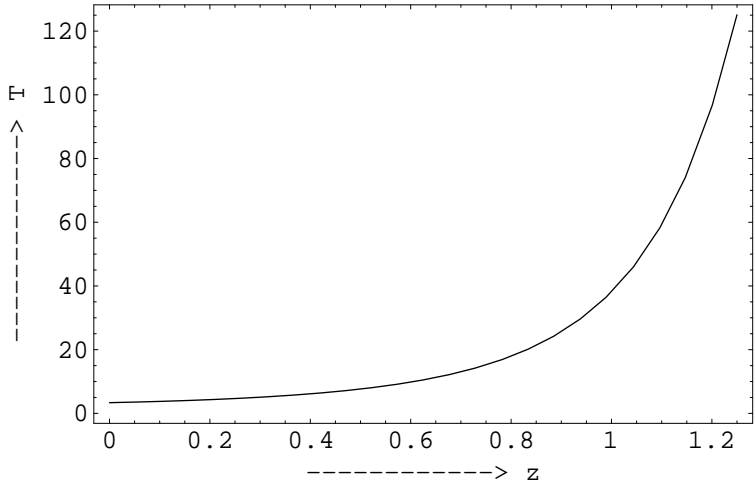,height=65mm,width=100mm}}
\caption{\normalsize{\em Plot of neutrino temparature $T$ ( in the units 
$T_{0} = 1$ ) vs. redshift $z$ for $\eta = \eta_{0}\rho^{1/2}$ and 
$\tau = \frac{\eta}{\rho}$.}}
\label{fig3}
\end{figure}
\section{Conclusion :~}
As a conclusion, one can say that a sufficient bulk viscous stress in the 
neutrino distribution can potentially serve the 
purpose of a dark energy. This does 
not require any ill-motivated scalar potential or an otherwise unwarranted 
modification of general relativity. However, the model is far from 
being complete. The amount of the bulk viscous stress has to be sufficient 
to drive the acceleration, and the correct phenomenological connection 
between the quantities like $\eta$, $\tau$ and $\rho$ has to be found 
out so that temperature of neutrinoes has a realistic profile. The neutrino 
viscosity can in fact give a wide range of accelerating models. One of them, 
the effective $\Lambda$CDM model is discussed here. But the other example 
mentioned also works, and leaves a possibility of finding many others 
within the scope of it. It has to be searched which solution is favoured 
from the consideration of stability as well as that of observational bounds. 
\section{Acknowledgement :~}
Authors would like to thank Relativity and Cosmology Research Centre at 
Jadavpur University, Kolkata where part of this work has been done.


\begin{thebibliography}{25}
\bibitem{1}A. G. Riess et al, Astron. J. {\bf 116}, 1009 (1998);\\
          S. Perlmutter et al, Bull. Am. Astron. Soc. {\bf 29}, 1351 (1997);\\
          S. Perlmutter et al, Astrophys. J. {\bf 517}, 565 (1999);\\
           J. L. Tonry et al, Astrophys. J. {\bf 594}, 1 (2003),  
                            [astro-ph/0305008].
\bibitem{2}S. Bridle, O. Lahav, J. P. Ostriker and P. J. Steinhardt,
                  Science {\bf 299}, 1532 (2003);\\
           C. Bennet et al., Astrophys. J. Suppl. {\bf 48}, 1 (2003),  
                                  [astro-ph/0302207];\\
           G. Hinshaw et al., Astrophys. J. Suppl. {\bf 148}, 135 (2003), 
                              [astro-ph/0302217];\\
           A. Kogut et al., Astrophys. J. Suppl. {\bf 148}, 161 (2003), 
                              [astro-ph/0302213];\\
           D. N. Spergel et al., Astrophys. J. Suppl. {\bf 148}, 175 (2003),
                               [astro-ph/0302209].
\bibitem{3}V. Sahni, A. A. Starobinski, Int. J. Mod. Phys. D 
                  {\bf 9}, 373 (2000);\\
           T. Padmanabhan, Phys. Rept. {\bf 380}, 235 (2003), 
                               [hep-th/0212290].
\bibitem{4}J. Martin, astro-ph/0803.4076.
\bibitem{5}N. Banerjee and D. Pavon, Phys. Rev. D {\bf 63}, 043504 (2001).
\bibitem{6}N. Banerjee and S. Das, Mod. Phys. Lett. A {\bf 21}, 2663 (2006);\\
           S. Das and N. Banerjee, Gen. Relativ. Gravit. 
                {\bf 38}, 785 (2006).
\bibitem{7}D. Pavon and B. Wang, gr-gc/0712.0565. 
\bibitem{8}M. C. Bento, O. Bertolami and A. A. Sen, Phys. Rev. D {\bf 66}, 
                043507 (2002),[gr-qc/0202064];\\
           M. C. Bento, O. Bertolami and A. A. Sen, Phys. Lett. B {\bf 575}, 
                172 (2003), [astro-ph/0303538]. 
\bibitem{9}S. Capozziello, S. Carloni, A. Troisi, astro-ph/0303041;\\
           S. Capozziello, V. F. Cardone, S. Carloni, A. Troisi, 
                     Int. J. Mod. Phys. D {\bf 12}, 1969 (2003),  
                       [astro-ph/0307018];\\
           S. M. Carroll, V. Duvvuri, M. Trodden, M. S. Turner, 
                     Phys. Rev. D {\bf 70}, 043528 (2004), [astro-ph/0306438];\\
           D. N. Vollick, Phys. Rev. D {\bf 68}, 063510 (2003);\\
           S. Carloni, P.K.S. Dunsby, S. Capozziello and A. Troisi,
                     Class. Quantum. Grav. {\bf 22}, 4839 (2005),  
                  [gr-qc/0410046];\\
           S. Nojiri and S. D. Odintsov, Phys. Rev. D 
                      {\bf 68}, 123512 (2003);\\
           A. Borowiec and M. Francaviglia, Phys. Rev. D {\bf 70}, 
                  043524 (2004), [hep-th/0403264];\\
           S. Nojiri and S. D. Odintsov, Gen. Relativ. Gravit. {\bf 36}, 
             1765 (2003), [hep-th/0308176];\\
           S. Das, N. Benerjee and N. Dadhich, Class. Quant. Grav. {\bf 23}, 
                 4159 (2006), [astro-ph/0505096]. 
\bibitem{10}V. Sahni, astro-ph/0403324;\\
            E. J. Copeland, M. Sami and S. Tsujikawa, Int. J. Mod. Phys. D 
                  {\bf 15}, 1753 (2006), [hep-th/0603057];\\
            T. Padmanabhan, astro-ph/0602117.
\bibitem{11}E. Ma and U. Sarkar, Phys. Lett. B {\bf 638}, 356 (2006), 
              [hep-ph/0602116];\\
            P. Gu, H. He and U. Sarkar, Phys. Lett. B {\bf 653}, 
                419 (2007), [hep-th/0704.2020];\\
            A. W. Brookfield, C. van de Bruck, D. F. Mota and 
            D. Tocchini - Valentini, Phys. Rev. Lett. {\bf 96}, 061301 (2006),
               [astro-ph/0503349];\\
            A. W. Brookfield, C. van de Bruck, D. F. Mota and 
            D. Tocchini - Valentini, Phys, Rev. D {\bf 73}, 083515 (2006), 
               [astro-ph/0512367];\\
            K. Ichiki and Y. - Y. Keum, astro-ph/0803.3142;\\
            K. Ichiki and Y. - Y. Keum, hep-ph/0803.2274;\\
            O. E. Bjaelde and S. Hannestad, astro-ph/0806.2146. 
\bibitem{12}U. Sarkar : {\it Particle and Astroparticle Physics}, (Taylor and 
            Francis, NewYork, London), 2008.
\bibitem{dolgov}A. D. Dolgov, hep-ph/0803.3887.
\bibitem{13}P. Coles and F. Lucchin : {\it  Cosmology: The Origin and 
            evolution of cosmic structure}, (Chichester, UK: Wiley), 2002.
\bibitem{14}B. Modak, Pramana {\bf 23}, 809 (1984).
\bibitem{15}C. J. Calkoen and S. R. De Groot, Phys. Lett. A 
               {\bf 83}, 319 (1981);\\
            N. Caderni and R. Fabbri, Phys. Rev. D {\bf 20}, 1251 (1979);\\
            N. Caderni and R. Fabbri, Phys. Lett. A {\bf 67}, 19 (1978).    
\bibitem{16}R. F. Sawyer, Phys. Rev. D {\bf 74}, 043527 (2006), 
               [astro-ph/0601525].
\bibitem{17}R. Maartens, astro-ph/9609119.
\bibitem{18}D. Pavon, J. Bafaluy and  D. Jou, Class. Quant. Grav. {\bf 8}, 
               347 (1991);\\
            A. Lindblom, Ann. Phys. (NY) {\bf 151}, 466 (1983);\\
            R. Maartens, Class. Quant. Grav. {\bf 12}, 1455 (1995);\\ 
            N. Banerjee and S. Sen, Phys. Rev. D {\bf 57}, 4614 (1998);\\
            V. Romano and D. Pavon, Phys. Rev. D {\bf 47}, 1396 (1993);\\
            V. Romano and D. Pavon, Phys. Rev. D {\bf 50}, 2572 (1994);\\
            N. Banerjee and A. Beesham, Pramana, {\bf 46}, 213 (1996).  
\bibitem{19}S. Das and N. Banerjee, Gen. Relativ. Gravit., 
                {\bf 37}, 1695 (2005). 
\end{thebibliography}
\end{document}